\documentclass[twocolumn,aps,showpacs]{revtex4}
\usepackage[latin9]{inputenc}
\usepackage{amsmath}
\usepackage{amssymb}
\usepackage{graphicx}

\makeatletter

\providecommand{\tabularnewline}{\\}

\@ifundefined{textcolor}{}
{%
 \definecolor{BLACK}{gray}{0}
 \definecolor{WHITE}{gray}{1}
 \definecolor{RED}{rgb}{1,0,0}
 \definecolor{GREEN}{rgb}{0,1,0}
 \definecolor{BLUE}{rgb}{0,0,1}
 \definecolor{CYAN}{cmyk}{1,0,0,0}
 \definecolor{MAGENTA}{cmyk}{0,1,0,0}
 \definecolor{YELLOW}{cmyk}{0,0,1,0}
}

\providecommand{\tabularnewline}{\\}

\@ifundefined{textcolor}{}{%
 \definecolor{BLACK}{gray}{0}
 \definecolor{WHITE}{gray}{1}
 \definecolor{RED}{rgb}{1,0,0}
 \definecolor{GREEN}{rgb}{0,1,0}
 \definecolor{BLUE}{rgb}{0,0,1}
 \definecolor{CYAN}{cmyk}{1,0,0,0}
 \definecolor{MAGENTA}{cmyk}{0,1,0,0}
 \definecolor{YELLOW}{cmyk}{0,0,1,0}
}

\usepackage{color}

\def\NOT(#1,#2){\OneQubitGate(#1,#2){$X$}}

\makeatother

\begin{document}

\title{Pulse sequences for controlled 2- and 3-qubit gates in a hybrid quantum
register}

\author{Jingfu Zhang, Swathi S Hegde and Dieter Suter\\
 Fakultaet Physik, Technische Universitaet Dortmund,\\
 D-44221 Dortmund, Germany }

\date{\today}
\begin{abstract}
We propose and demonstrate a quantum control scheme for hybrid quantum
registers that can reduce the operation time, and therefore the effects
of relaxation, compared to existing implementations. It combines resonant
excitation pulses with periods of free precession under the internal
Hamiltonian of the qubit system. We use this scheme to implement quantum
gates like controlled-NOT operations on electronic and nuclear spins
of the nitrogen-vacancy center in diamond. As a specific application,
we transfer population between electronic and nuclear spin qubits
and use it to measure the Rabi oscillations of a nuclear spin in a
system with multiple coupled spins. 
\end{abstract}

\maketitle

\section{Introduction}

High precision quantum control is required in various fields,
such as quantum computing \cite{QCNJP10,nielsen,Stolze:2008xy}. The
gate operations used for quantum control often rely on resonant electromagnetic
fields that drive the targeted qubits near a resonant transition.
While this drive operation should be strong, to dominate over unwanted
effects and to allow short gate operations, the strength of the control
field is often also limited by the requirement that it must not affect
qubits that are not targeted in the specific operation.

Meeting these requirements becomes progressively more challenging
as the number of qubits increases, as larger systems have more resonant
transitions that must be considered. In many systems, the strengths
of the couplings between the qubits, which are essential for multi-qubit
gates, cover a significant range of values. If the couplings are weak,
the duration of the gate operations that rely on these interactions,
increases correspondingly. Multi-qubit operations often rely on transition
selective pulses \cite{quant-ph:9809045,PhysRevLett.93.130501,PhysRevLett.115.110502,nature17404,nature10401},
which are designed to drive only a single transition with a resonant
field. The condition that this field should be weak compared to the
strength of the coupling constant may then be in conflict with the
requirement that the gate operation should be fast compared to the
relaxation time.

To avoid this conflict, we propose and demonstrate here an alternative
approach, which is often used in liquid-state magnetic resonance but
less frequently in solid-state systems like the diamond nitrogen-vacancy (NV) center
\cite{Suter201750}. It is based on combining hard pulses, i.e. driving
fields that are strong compared to the couplings, with periods of
free precession, where the coupling differentiates between the different
qubit states. The resulting gate durations are the minimum possible
for the given system. Compared to the approach with transition-selective
pulses, this technique also offers the possibility to combine the
pulses with techniques for reducing decoherence \cite{RevModPhys.88.041001},
such as dynamical decoupling. In particular, dynamical decoupling pulses can be
combined with gate operations for designing gates that are protected
against environmental noise \cite{PhysRevLett.112.050502, PhysRevLett.115.110502}.

\section{Electron spin + $^{14}$N nuclear spin}

\label{CNOTen}

\begin{figure}
\centering{}\includegraphics[width=1\columnwidth]{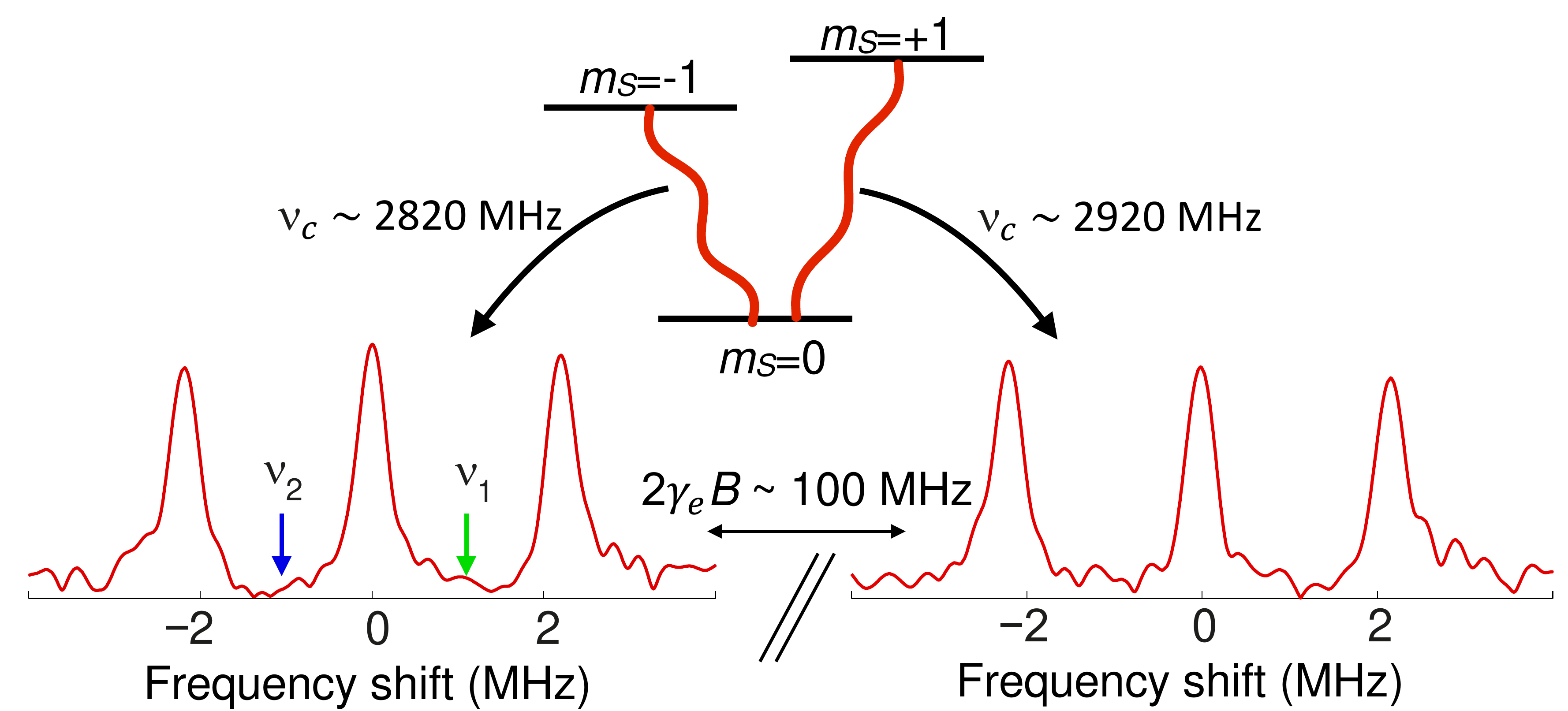} 
\caption{(color online). Energy level system and spectra of the ESR transitions between the
states with $m_{S}=0$ and $\mp1$, obtained as Fourier-transforms
of time-domain signals. The origins of the frequency axes are set
to $D\pm\gamma_{e}B$. The vertical arrows indicate the carrier frequencies
$\nu_{1,2}$ for the MW pulses to implement the conditional operations
$U_{1}$ and $U_{2}$, respectively. \label{figFFT}}
\end{figure}

\subsection{Hamiltonian}

The system that we consider consists of the electron spin and the
$^{14}$N nuclear spin system of a single NV center. If the magnetic field is oriented along the NV symmetry axis,
the relevant Hamiltonian can be written as ~\cite{PhysRevB.89.205202,Suter201750}
\begin{equation}
\frac{1}{2\pi}\mathcal{H}_{E,N}=DS_{z}^{2}-\gamma_{e}BS_{z}+PI_{z}^{2}-\gamma_{n}BI_{z}+AS_{z}I_{z}.\label{Hamsim}
\end{equation}
Here $S_{z}$ and $I_{z}$ are $z$-components of the spin-1 operators
for the electronic and nuclear spins, respectively. The zero-field
splitting is $D=2.87$ GHz, the nuclear quadrupolar splitting is $P=-4.95$
MHz, and the hyperfine coupling $A=-2.16$ MHz ~\cite{PhysRevB.89.205202,PhysRevB.47.8816,Yavkin16}.
The electronic gyromagnetic ratio is $\gamma_{e}=-28$ GHz/T, and
the nuclear gyromagnetic ratio $\gamma_{n}=3.1$ MHz/T. In the experiments,
the static field strength is about $1.8$ mT, which results in a separation
of the two electron spin resonance (ESR) transitions by about $100$
MHz. Figure \ref{figFFT} shows the spectra of the ESR transitions
between the states with $m_{S}=0$ and $\mp1$, obtained in Ramsey-type
free-induction decay (FID) time-resolved experiments, using resonant
microwave pulses with Rabi frequencies of about 10 MHz for excitation
and detection.

Since the experimental Rabi frequency is small compared to the separation
of the two ESR transitions, the individual experiments are confined
to a subspace of the full Hilbert space. We consider here the six-dimensional
subspace spanned by the states 
\begin{equation}
\{|0\rangle_{e},|-1\rangle_{e}\}\otimes\{|1\rangle_{n},|0\rangle_{n},|-1\rangle_{n}\},\label{space6DFG}
\end{equation}
which is associated to the ESR transition between $|0\rangle_{e}$
and $|-1\rangle_{e}$, with a transition frequency $D+\gamma_{e}B\sim2820$
MHz. It thus contains one qubit (electron spin; target qubit) and
one qutrit ($^{14}$N spin; control qutrit). The relevant Hamiltonian
is then 
\begin{equation}
\frac{1}{2\pi}\mathcal{H}_{E,N}^{eff}=\frac{\nu}{2}\sigma_{z}\otimes E_{3}+\frac{A}{2}\sigma_{z}\otimes I_{z},\label{Ham62eff}
\end{equation}
written in the rotating frame at the carrier frequency $\nu_{c}$
of the microwave, where $\sigma_{z}$ denotes the $z$- component
of the Pauli matrix for the pseudo spin 1/2 of the electron spin in
the space $\{|0\rangle_{e},|-1\rangle_{e}\}$, $E_{3}$ the identity
operator in 3 dimensions, and 
\begin{eqnarray}
\nu=(D+\gamma_{e}B)-\nu_{c}\label{freqshift}
\end{eqnarray}
is the effective transition frequency of the electron spin qubit in
the rotating frame. In addition, the interaction representation also
eliminates the quadrupole and Zeeman interactions of $^{14}$N nuclear
spin, which are irrelevant for the purpose of this work.

\subsection{Unitary gate operations}

The basic microwave (MW) pulse sequence for a controlled operation
consists of two $\pi/2$ pulses with a $\pi/2$ relative phase shift,
separated by a period $\tau$ of free precession, during which the
hyperfine interaction causes differential precession, depending on
the state of the nuclear spin. This pulse sequence is closely analogous
to sequences used in nuclear magnetic resonance (NMR) quantum computing
for implementing CNOT gates \cite{1019} and in electron-nuclear double
resonance (ENDOR) for polarizing the nuclear spin \cite{Yavkin16}.
In the above mentioned reference frame, the unitary operation generated
by the pulse sequence is 
\begin{eqnarray}
U= & - & \frac{i}{2\sqrt{2}}[(c_{-}(\sigma_{z}-iE_{2})+c_{+}(\sigma_{x}+\sigma_{y})]\nonumber \\
 &  & \otimes[\sin(\pi A\tau)]I_{z}\nonumber \\
 & + & \frac{i}{2\sqrt{2}}[c_{+}(\sigma_{z}-iE_{2})-c_{-}(\sigma_{x}+\sigma_{y})]\nonumber \\
 &  & \otimes\{[\cos(\pi A\tau)-1]I_{z}^{2}+E_{3}\},\label{UCNOTt}
\end{eqnarray}
with $c_{\pm}=\cos(\pi\nu\tau\pm\frac{\pi}{4})$. To obtain Eq. (\ref{UCNOTt}),
we used 
\begin{equation}
e^{-i\theta\sigma_{z}I_{z}}=-i\sin\theta\sigma_{z}I_{z}+[\cos\theta-1]E_{2}\otimes I_{z}^{2}+E_{2}\otimes E_{3}\label{coupevl}
\end{equation}
and assumed that the MW pulses are ideal, with duration zero.

When $\tau=1/(2|A|)$, $U$ corresponds to conditional qubit-qutrit
operations, e.g. $U_{1}=U(\nu_{c}\to\nu_{1}=D+\gamma_{e}B-A/2)$ and
$U_{2}=U(\nu_{c}\to\nu_{2}=D+\gamma_{e}B+A/2)$. Table \ref{tablequtrit}
summarizes the effects of $U_{1}$ and $U_{2}$ on the basis states
of the system. Up to some phase factors, they represent Controlled-NOT
(CNOT) gates.

\begin{table}
\begin{tabular}{|c|c|c|}
\hline 
Input  & Output of $U_{1}$  & Output of $U_{2}$\tabularnewline
\hline 
0 1  & {\color{blue} -1 1}  & 0 1 \tabularnewline
\hline 
0 0  & 0 0  & {\color{green} -1 0} \tabularnewline
\hline 
0 -1  & {\color{blue}-1 -1}  & 0 -1\tabularnewline
\hline 
-1 1  & {\color{blue}0 1}  & -1 1\tabularnewline
\hline 
-1 0  & -1 0  & {\color{green}0 0} \tabularnewline
\hline 
-1 -1  & {\color{blue}0 -1}  & -1 -1 \tabularnewline
\hline 
\end{tabular}\caption{(color online). Output states $m_{S}$ and $m_{I}$ of $U_{1}$ and
$U_{2}$ for the system consisting of one qubit (electron $S$; target)
and one qutrit (nuclear spin $I$; control). global phase factors
have been ignored. The states that are unchanged are shown in black
in the results, the transformed states in blue (for $U_{1}$) and
green (for $U_{2}$). }
\label{tablequtrit} 
\end{table}

\subsection{Experimental implementation}

\label{exphalf}

\begin{figure}
\centering{}\includegraphics[width=1\columnwidth]{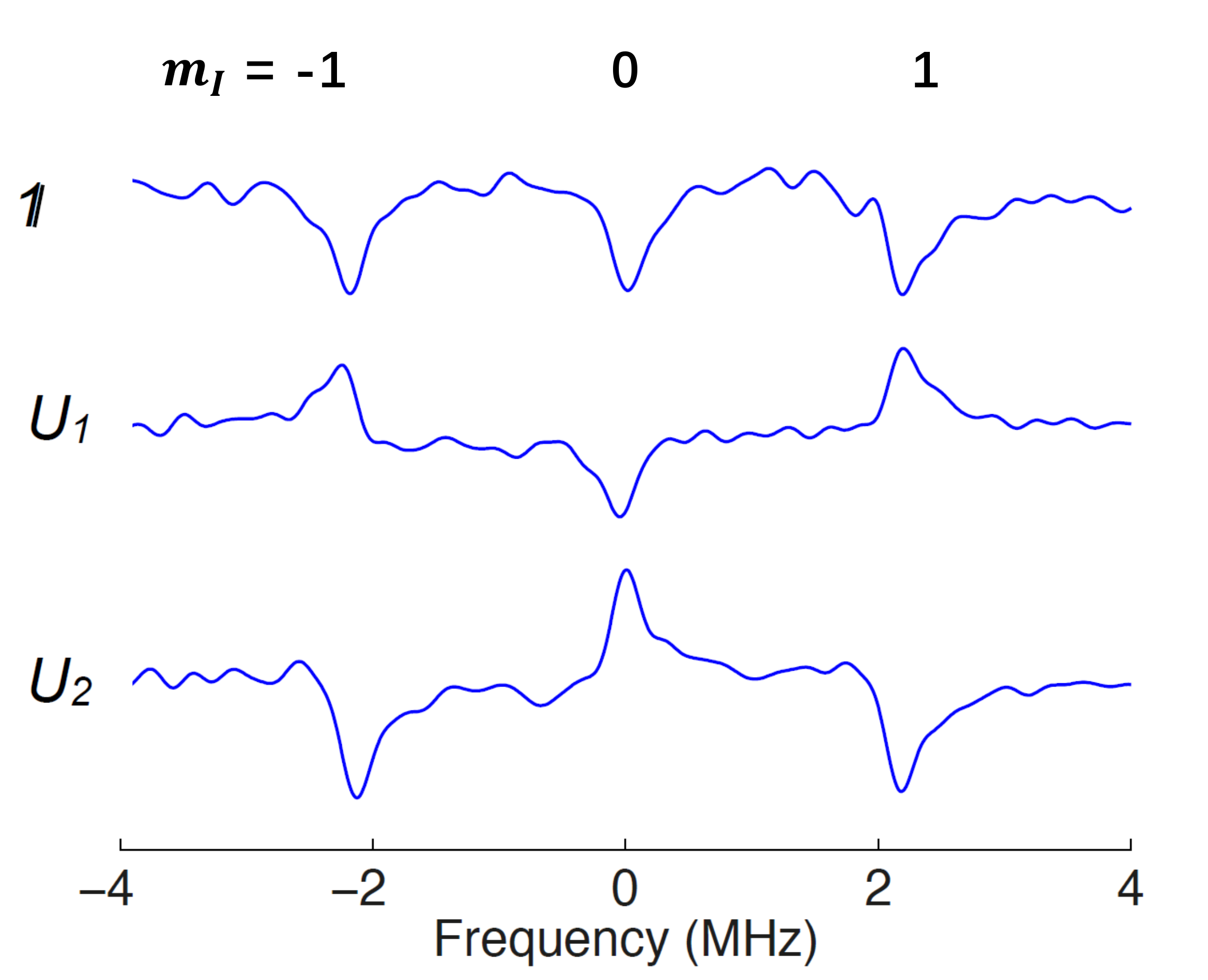} 
\caption{(color online). Experimental ESR spectra obtained as Fourier-transforms of the FIDs,
after applying the identity (top), $U_{1}$ (middle) and $U_{2}$
(bottom) gate operations. The quantum numbers above the spectra indicate
the states of $^{14}$N corresponding to the peaks. \label{fig:Experimental-ESR-spectra}}
\end{figure}

The experiments were performed at room temperature, using a diamond
sample with natural abundance (1.1 \% $^{13}$C). The transverse relaxation
time $T_{2}^{*}$ of the electron spin was $\approx2.5$ $\mu$s,
measured in a Ramsey-type FID experiment. To test the sequence, we
first initialized the electron spin of the NV center system into the
$m_{S}=|0\rangle$ state by a laser pulse. To a first approximation,
the nuclear spin is not affected by the laser pulse but remains unpolarized.
Therefore the state of the two spins after the laser pulse is 
\begin{equation}
\rho_{ini}=|0\rangle\langle0|\otimes\frac{E_{3}}{3},\label{initials2}
\end{equation}
where $E_{3}$ denotes the $3\times3$ unit operator. After the MW
pulses, a second laser pulse measures the population of the $m_{S}=|0\rangle$
state. Figure \ref{fig:Experimental-ESR-spectra} shows the experimental
ESR spectra obtained as the Fourier-transforms of the FID. The uppermost
trace shows the normal spectrum, i.e., without any gate operation;
the second and third trace were obtained after applying the gates
$U_{1}$ and $U_{2}$. The signs of the peaks show that the electron
spin states were inverted if the $^{14}$N nuclear spin was in the
$m_{I}=\pm1$ state (for $U_{1}$) or in the $m_{I}=0$ state (for
$U_{2}$), as expected from table \ref{tablequtrit}.

\section{Electron, $^{14}$N and $^{13}$C}

\label{sec3spins}

\begin{figure}
\centering{}\includegraphics[width=1\columnwidth]{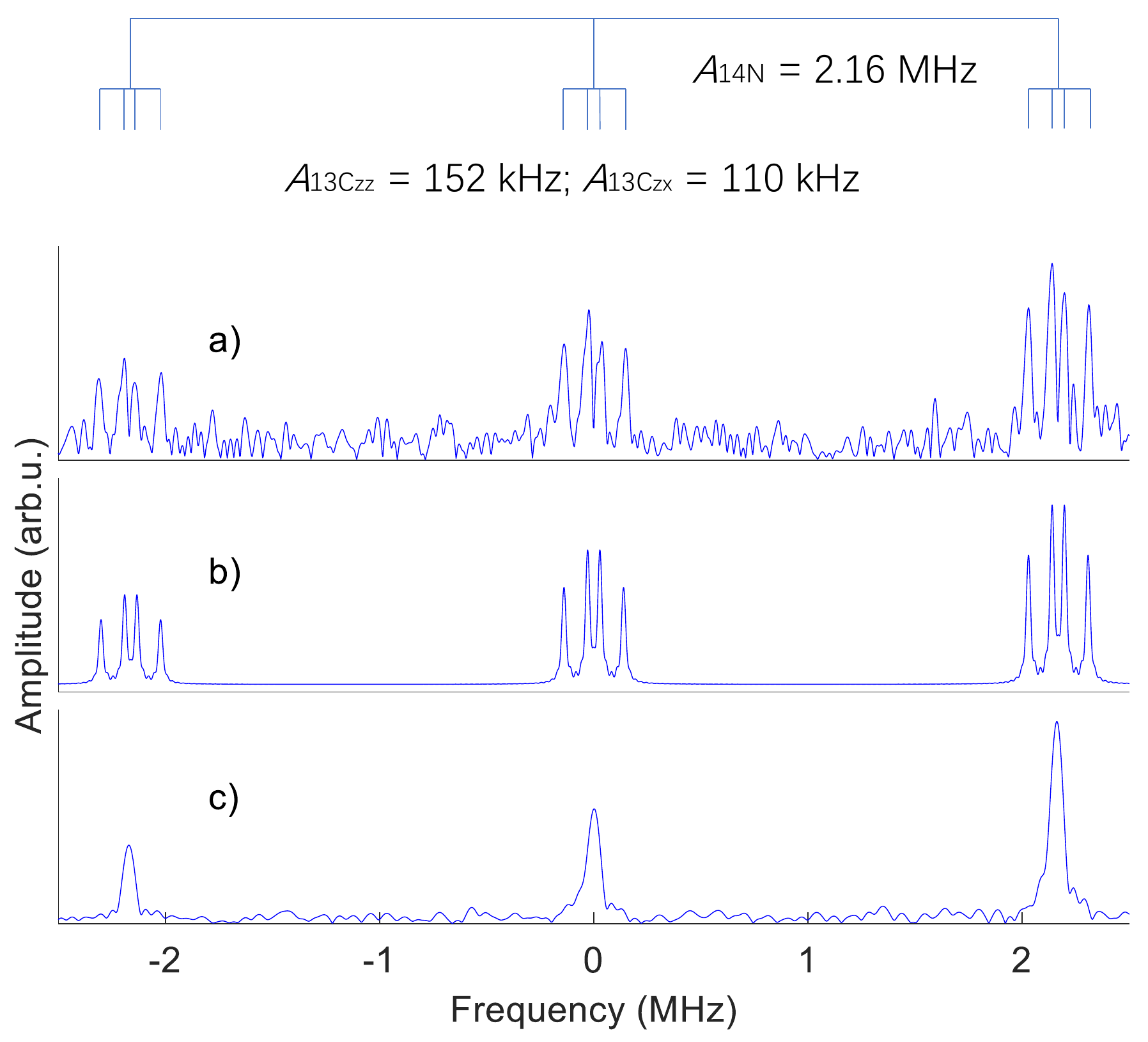}
\caption{(color online). Spectra of the electron spin coupled to one $^{14}$N and one $^{13}$C
nucleus, obtained from an FID experiment (a) and the corresponding
simulation (b). The origin of the frequency axis is $\nu_{1}=D+\gamma_{e}B$.
Due to the hyperfine coupling with the $^{13}$C nucleus, each peak
of the electron-$^{14}$N system is split into four, shown by the
schematic diagram at the top. The spectrum (c) was obtained from another
electron spin which is only coupled to a $^{14}$N .}
\label{figFFT3} 
\end{figure}

As one example of the conditional gate operation $U_{2}$, we use
it for selective population transfer between the electronic and nuclear
spins in a three-spin system consisting of the electron, the $^{14}$N
and one $^{13}$C nuclear spin. Figure \ref{figFFT3} (a) shows the
spectrum of the electron spin, obtained from an FID experiment. The
interaction with the $^{13}$C nuclear spin splits each resonance
line of the electron-$^{14}$N spin system into four lines. This indicates
that the strength of the hyperfine interaction of the $^{13}$C is
comparable to its nuclear Zeeman interaction, which is 165 kHz under
our experimental conditions. Figure \ref{figFFT3} (b) shows the simulated
spectrum, with the couplings shown in the panel at the top of the
figure.

\begin{figure}
\centering{}\includegraphics[width=1\columnwidth]{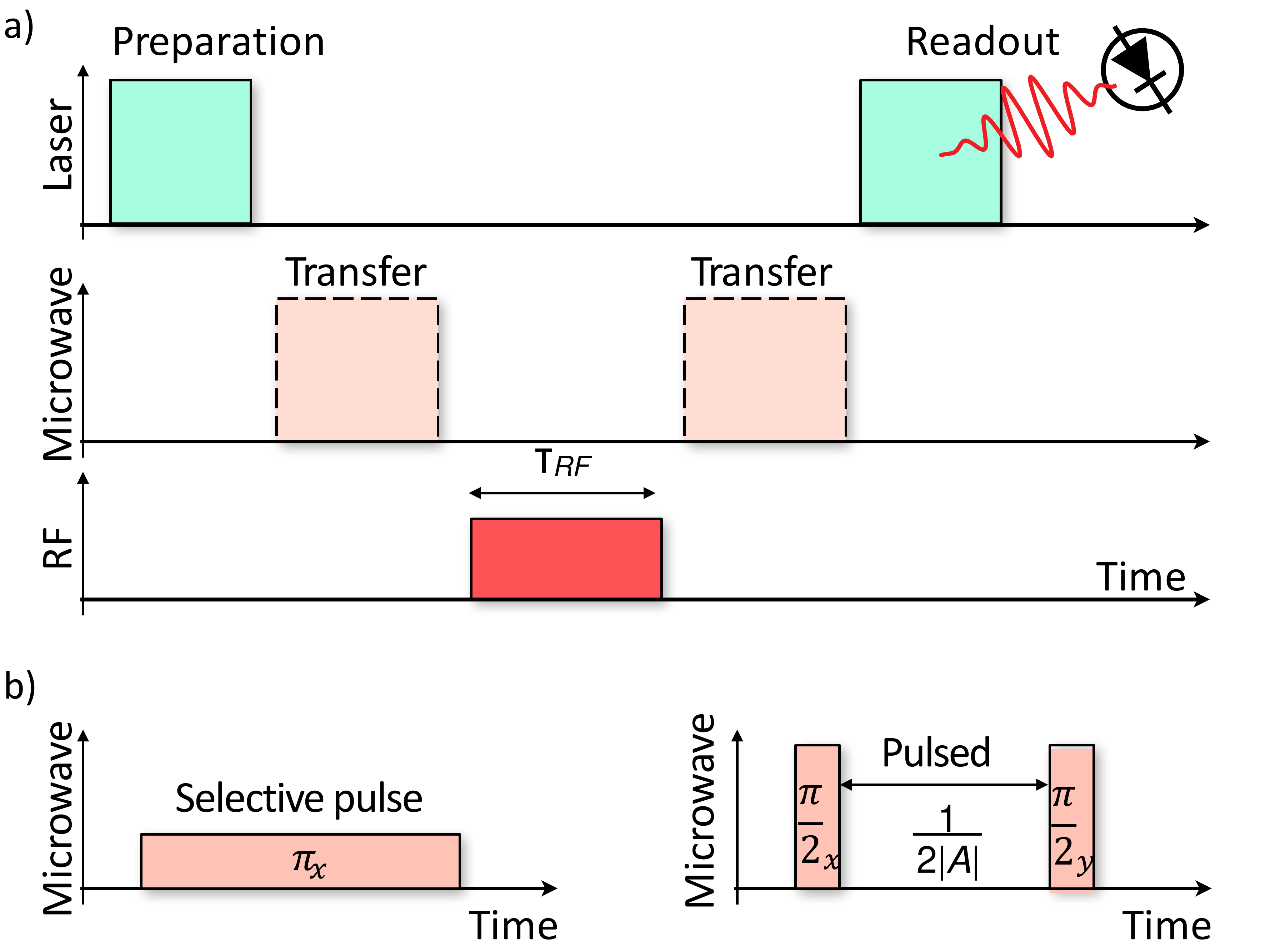} 
\caption{(color online). (a) Pulse sequence for measuring Rabi frequency of the nuclear spin.
The carrier frequency of the RF pulse is resonant with an NMR transition.
The first transfer operation polarizes the nuclear spin, and the second
acts as a readout to measure the remaining population after the RF
pulse. (b) The MW pulse used for the population transfer in previous
works (left) \cite{PhysRevA.87.012301,dobrovitski2012}, and in the
present work (right). In the selective pulse method, the Rabi frequency
of the MW pulse is about 0.2 MHz, while in pulsed transfer, the pulses
are hard pulses with Rabi frequencies of about 10 MHz. \label{figCNOTe}}
\end{figure}

An effective population transfer requires that the $^{14}$N nuclear
spin controls the evolution of the electron spin, with little perturbation
from the $^{13}$C spin. In our experiment, the coupling from the
passive spin is not negligible, in contrast to previous experiments,
where there was no passive spin \cite{dobrovitski2012}, or the coupling
from the passive spin was negligibly small compared to the active
spin \cite{PhysRevA.87.012301}. We first use the population transfer
to polarise the $^{14}$N nuclear spin, as shown in Figure \ref{figCNOTe}.
After the transfer, we measure the Rabi frequency of the $^{14}$N
spin by applying a constant radio-frequency (RF) field of variable duration $\tau_{RF}$.
After the RF pulse, we use another conditional gate operation for
transferring the remaining nuclear spin polarisation back to the electron
spin for detection.

The experiments were implemented in a $^{12}$C enriched diamond sample,
where decoherence due to $^{13}$C nuclear spins is small and the
coherence time of the electron spin is $\approx$ 20 $\mu$s. To evaluate
the effect of the passive $^{13}$C, we compare two centers, one with
and one without the $^{13}$C. The NV axes of the two centers point
in the same direction, and the NMR transition frequencies in the concerned
subspace $\{m_{S}=|0\rangle,|-1\rangle$ \} for these two centers
were 4.981, 4.905, 2.822, and 7.075 MHz, for the transitions between
the states $|0,0\rangle\leftrightarrow|0,-1\rangle$, $|0,0\rangle\leftrightarrow|0,1\rangle$,
$|-1,0\rangle\leftrightarrow|-1,-1\rangle$ and $|-1,0\rangle\leftrightarrow|-1,1\rangle$.
Figure \ref{figFFT3} (c) shows the spectrum obtained from the electron
spin without a coupled $^{13}$C. For both centers, we performed measurements
with the two different transfer techniques.

\begin{figure}
\centering{}\includegraphics[width=1\columnwidth]{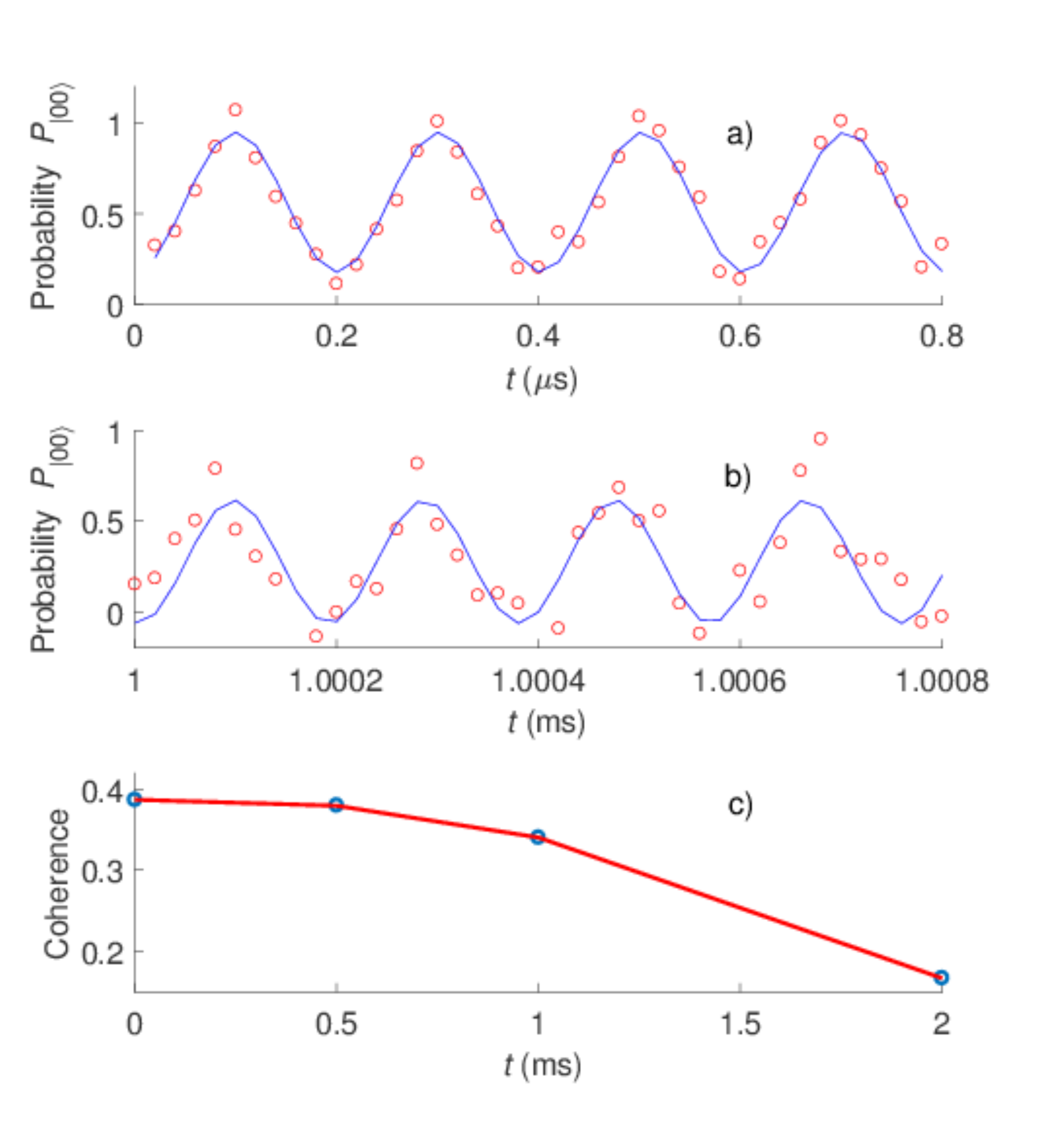}
\caption{(color online). Experiment results in estimating $T_{2}^{*}$ time for $^{14}$N.
(a-b) The probability of state $|0,0\rangle$ measured in the FID
experiments in various time period, where the measured data and fitting
results are indicated by circles and solid curves respectively. (c)
The coherence extracted from the FID experiments changes against time.
By fitting the measured data shown as the circles using a function
as $A_{0}e^{-(t/T_{2}^{*})^{k}}$, we can obtain $T_{2}^{*}=2.1$
ms, $A_{0}=0.39$ and $k=2.7$. \label{resFIDN}}
\end{figure}

For estimating the $T_{2}^{*}$ for $^{14}$N, we
used the standard pulse sequence to measure the FID of the $^{14}$N
\cite{dobrovitski2012,PhysRevA.87.012301}. In the measurement, the
MW and RF pulses were transition-selective, resonant with the transitions
$|0,0\rangle\leftrightarrow|-1,0\rangle$ and $|0,0\rangle\leftrightarrow|0,-1\rangle$,
respectively. We performed the experiments by choosing various time
periods, and measuring the population of the state $|0,0\rangle$.
Figure \ref{resFIDN} shows the experimental results. Through fitting
the coherence extracted from the series FID experiments, we can obtain
$T_{2}^{*}=2.1$ ms. Since this value is close to the
longitudinal relaxation time $T_{1}$ of the electron spin, which was measured
as $\approx3.5$ ms, the precision of this strategy is limited by
the $T_{1}$ value of the electron.

\begin{figure}
\centering{}\includegraphics[width=1\columnwidth]{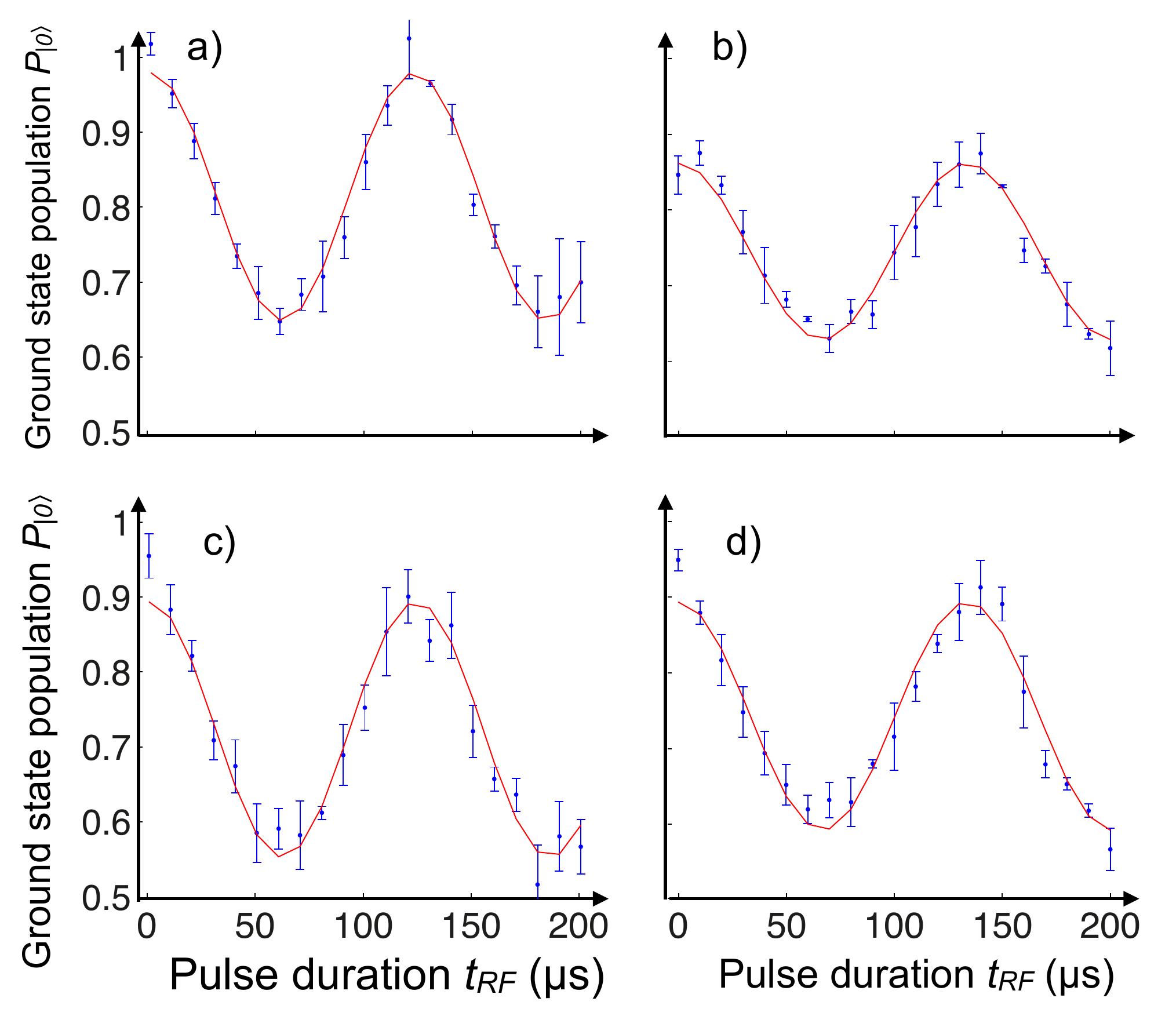}
\caption{(color online). Results of $^{14}$N Rabi experiments in the NV centers without and
with coupled $^{13}$C, shown as left and right columns, from previous
(top) and the new sequence proposed here (bottom). \label{RabiNewres}}
\end{figure}

We firstly applied the pulse sequence used in the previous works to
a center without coupled $^{13}$C. The Rabi frequency of the MW pulses
was $0.19$ MHz, with the carrier frequency set to the transition
between the $|0,0\rangle$ and $|-1,0\rangle$ states. The frequency
of the RF pulse was 4.981 MHz. The experimentally measured Rabi nutation
is shown in Figure \ref{RabiNewres} (a), together with a fit to the
function 
\begin{equation}
P_{|0\rangle}=\alpha+\beta\cos(2\pi\nu_{R}t_{RF}),\label{funcfit}
\end{equation}
where $\nu_{R}$ denoted the measured Rabi frequency for $^{14}$N,
and $\alpha$ and $\beta$ are two constant. Table \ref{tableNparams}
shows the parameter values obtained by fitting the experimental data.
Here we chose Figure \ref{RabiNewres} (a) as a reference for evaluating
the following experiments.

Figure \ref{RabiNewres} (b) shows the results obtained in the center
coupled with $^{13}$C, from the previous pulse sequence in Figure
\ref{figCNOTe}, where the Rabi frequency of the two MW pulses is
$0.21$ MHz. By comparing with Figure \ref{RabiNewres} (a), one can
find that the signal was degraded by the couplings from $^{13}$C,
shown as the loss of the strength of the signal by $12\%$ and the
decrease of the amplitude of the oscillation as $29\%$, since the
Rabi frequency of the MW pulses is not large enough compared with
the splitting caused by $^{13}$C, which is about $0.15$ MHz.

Figure \ref{RabiNewres} (c-d) shows the results obtained from our
modified sequence shown in Figure \ref{figCNOTe}, applied to the
centers without and with the coupled $^{13}$C. Table \ref{tableNparams}
shows the fitted parameters. Comparing the values for (d) and (b)
shows a significant advantages for the new pulse sequence, which increases
the oscillation amplitude of the signal by $21\%$. This advantages
can be traced to the reduced effect of the coupling to $^{13}$C due
to the faster operations. One can estimate that the infidelity caused
by the coupling from $^{13}$C can be reduced by $\sim A/\nu_{1,MW}>10$,
where $\nu_{1,MW}=0.21$ MHz is the Rabi frequency of the transition
selective MW $\pi$ pulses in Figure \ref{figCNOTe}.

\begin{table}
\begin{tabular}{|c|c|c|c|}
\hline 
 & $\alpha$  & $\beta$  & $\nu_{R}$ (kHz)\tabularnewline
\hline 
Fig. (a)  & $0.816$  & $0.164$  & 8.2 \tabularnewline
\hline 
Fig. (b)  & $0.743$  & $0.117$  & 7.5 \tabularnewline
\hline 
Fig. (c)  & $0.723$  & $0.168$  & 8.1\tabularnewline
\hline 
Fig. (d)  & $0.742$  & $0.151$  & 7.5\tabularnewline
\hline 
\end{tabular}\caption{Fit results from the experiment data shown in Figures \ref{RabiNewres}
(a-d). }
\label{tableNparams} 
\end{table}

\section{CCNOT gate }

\label{cnotC13}

$^{13}$C nuclear spins close to NV centers are interesting candidates
for qubits \cite{OpticsSpectroscopy91pp429}, provided effective gate
operations can be implemented. If the hyperfine coupling is strong,
2-qubit gates can be implemented by transition-selective pulses \cite{PhysRevA.87.012301}.
However, if multiple qubits are required, such as for the implementation
of quantum error correction, it also becomes important to control 
more remote $^{13}$C spins \cite{Nature204506,naturephoton}. In
these cases, selective pulses result in long gate times and as a result,
the coherence time of the electron spin limits the fidelity of the
overall gate operation. The pulsed scheme discussed above results
in significantly shorter gate times and can thus alleviate this problem.

In this section, we still consider a system consisting of one electron,
one $^{14}$N and and $^{13}$C spin. We implement CNOT in the electron-$^{13}$C
system, with the additional constraint that the $^{14}$N nuclear
spin is in the state $m_{I}=1$. In the three spin system, this operation
is a controlled-controlled NOT (CCNOT) gate, or Toffoli gate\cite{nielsen},
where $^{13}$C and $^{14}$N are the control spins, and the electron
spin is the target. Since our scheme to implement the CNOT relies
only on the secular component $A_{zz}$ of the hyperfine interaction,
we choose a center for which the quantization axis is close to the
NV-axis \cite{PhysRevLett.110.060502}. By aligning the field along
the NV- axis, we can approximate the relevant interaction Hamiltonian
between electron and $^{13}$C as 
\begin{equation}
\frac{1}{2\pi}\mathcal{H}_{c}=A_{zz}S_{z}s_{z},\label{HamcoupC13}
\end{equation}
where $s_{z}$ denotes the $z$-component of the $^{13}$C nuclear
spin operator.

\begin{figure}
\centering{}\includegraphics[width=1\columnwidth]{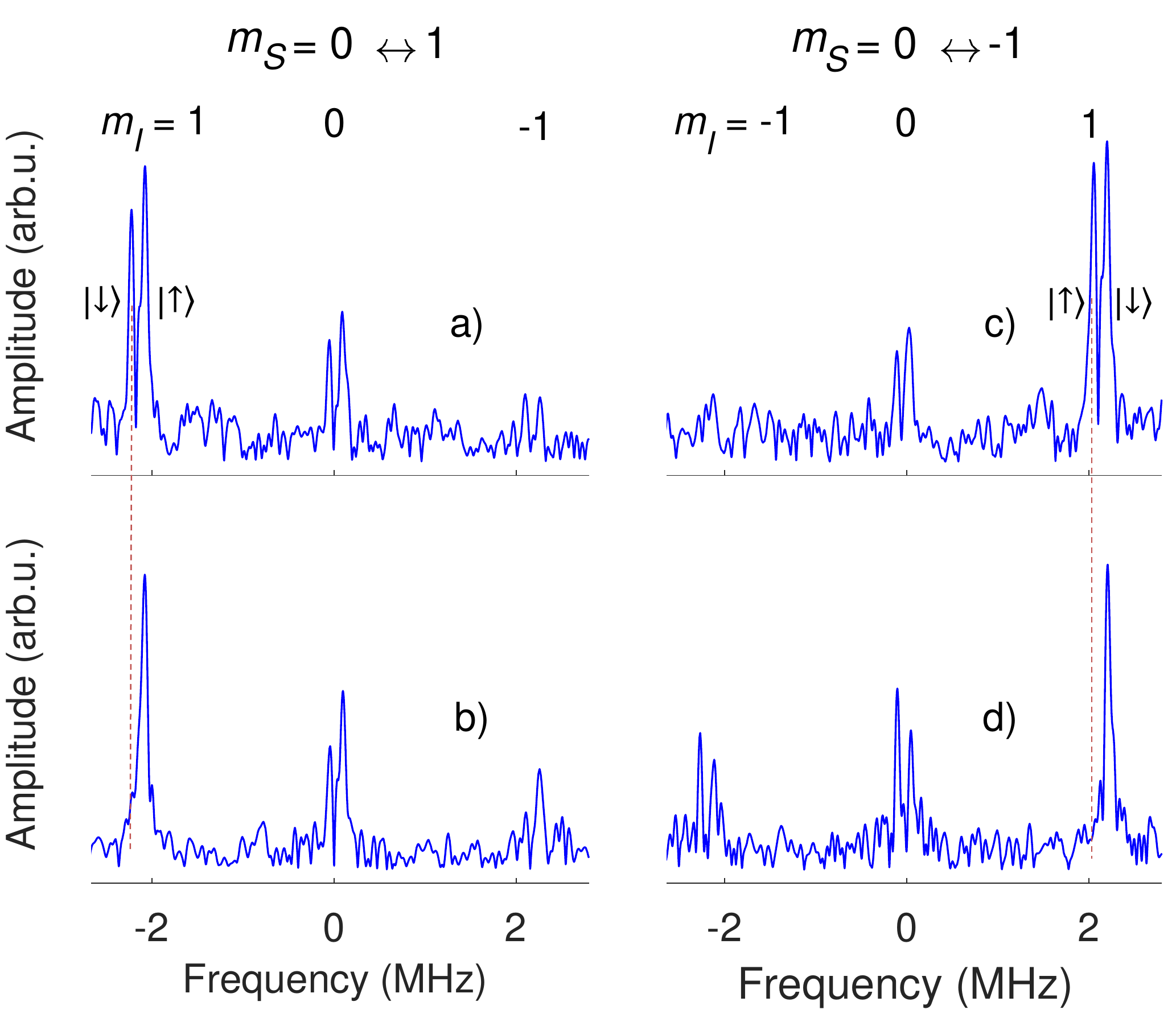}
\caption{(color online). Spectra of the electron spin coupled to one $^{14}$N and one $^{13}$C
nucleus. The spectra in (a) and (c) are obtained from FIDs using MW
pulses with the carrier frequencies set to the transition frequencies
between states $m_{S}=0$ and $\pm1$, and Rabi frequencies of 6 and
10 MHz, respectively. The spectra in (b) and (d) show the results
obtained from the FID after the CNOT gates were implemented in the
$m_{S}=0$ and $\mp1$ manifolds with $^{14}$N in state $m_{I}=1$,
respectively. The states of the $^{13}$C nuclear spin are indicated
by $\uparrow$, $\downarrow$. The origin of the frequency axis in
(a-b) is $D-\gamma_{e}B$, and in (c-d) $D+\gamma_{e}B$. The dashed
vertical lines indicate the positions of resonance lines where the
nuclear spins are in the state that corresponds to the control condition:
in this case, the gate operation flips the electron spin, which is
verified by the fact that the corresponding peaks are absent in the
spectra (b) and (d). \label{figCNOTC} }
\end{figure}

\begin{table}
\begin{tabular}{|c|c|c|}
\hline 
Input  & Output of CNOT$_{1}$  & Output of CNOT$_{2}$\tabularnewline
\hline 
1 $\uparrow$  & 1 $\uparrow$  & 0 $\uparrow$ \tabularnewline
\hline 
1 $\downarrow$  & 1 $\downarrow$  & 1 $\downarrow$ \tabularnewline
\hline 
0 $\uparrow$  & 0 $\uparrow$  & 1 $\uparrow$ \tabularnewline
\hline 
0 $\downarrow$  & -1 $\downarrow$  & 0 $\downarrow$\tabularnewline
\hline 
-1 $\uparrow$  & -1 $\uparrow$  & -1 $\uparrow$ \tabularnewline
\hline 
-1 $\downarrow$  & 0 $\downarrow$  & -1 $\downarrow$ \tabularnewline
\hline 
\end{tabular}\caption{ Output states electron (target) and $^{13}$C (control) spins of
the CNOT operations in the $m_{S}=0$ and $\mp1$ manifolds, indicated
as CNOT$_{1}$ and CNOT$_{2}$. }
\label{tablequbit} 
\end{table}

We still use the $^{12}$C enriched sample mentioned in the previous
section, and perform the experiments at room temperature. The time
$T_{2}^{*}$ is about 10 $\mu$s. Figure \ref{figCNOTC} (a-b) shows
the electron spin spectra obtained through FID measurements in the
$m_{S}=0$ and $m_{S}=\pm1$ manifolds, respectively. The number of
resonance lines indicates that in this center, the hyperfine tensor
component $A_{zx}$ is sufficiently small to be neglected (\textless{50}
kHz). The splitting of the peaks indicates that $A_{zz}\approx150$
kHz. In the following, we consider 2 subspaces of the full Hilbert
space, both of which correspond to two-qubit systems with one electron-spin
qubit and one $^{13}$C nuclear spin qubit, while the $^{14}$N nuclear
spin is in the $m_{N}=1$ state. The first subspace is spanned by
the states $m_{S}=0$ and $m_{S}=-1$ of the electron and the second
by the states $m_{S}=0$ and $m_{S}=1$. In these two subspaces, we
implement two slightly different CNOT gates, with the electron spin
as a target and the nuclear spin as the control qubit. In the first
subspace, we use the cotnrol condition that the $^{13}$C spin is
in the $\downarrow$ state, in the second subsystem that it is in
the state $\uparrow$. Table \ref{tablequbit} summarises these gates.

\begin{figure}
\centering{}\includegraphics[width=1\columnwidth]{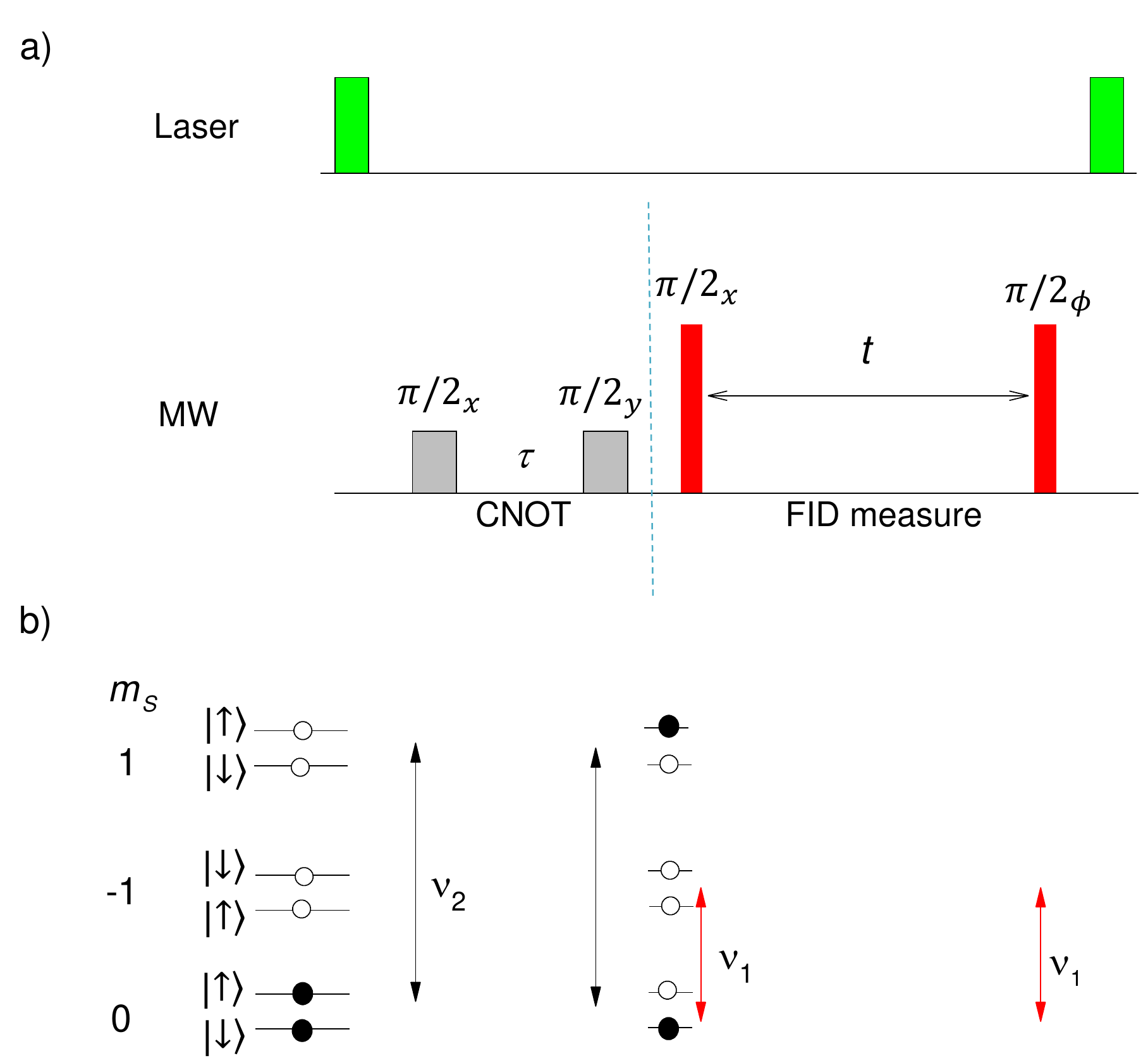}
\caption{(color online). (a) Pulse sequence for demonstrating the CNOT gate implemented in
the electron-$^{13}$C system. (b) Populations of the initial and
final states when the CNOT gate is applied to subspace with $m_{S}=0$$\leftrightarrow-1$,
$m_{I}=1$. We therefore do not consider the $^{14}$N states here.
The filled and empty circles denote population of 1/2 and 0, respectively.
The states marked by $|\uparrow\rangle$ and $|\downarrow\rangle$
are the $^{13}$C eigenstates. The carrier frequencies of the MW pulses
are set to the transition frequencies between $m_{S}=0$ and 1 or
-1, indicated as the lines with double arrows. $\nu_{1}$ and $\nu_{2}$
indicate the transition frequencies $D\pm\gamma_{e}B$, respectively.
The Rabi frequencies of the MW pulses are $0.5$ and $10$ MHz in
the CNOT gate and the FID measurement, respectively. \label{figCNOTCpulse}}
\end{figure}

Figure \ref{figCNOTCpulse} shows the pulse sequence used. The double
arrows indicate the carrier frequencies of the MW pulses. The Rabi
frequency of the pulses in the CNOT is $0.5$ MHz. These two pulses
are as hard pulses for the electron and $^{13}$C system, since the
Rabi frequency is much larger than the coupling to $^{13}$C, while
they are selective with respect to the state of the $^{14}$N. In
order to observe the effects of the CNOT gate, we measured the FID
signal. Figure \ref{figCNOTC} (c-d) shows the spectra of the electron
after the implementation of the CNOT gates in the $m_{S}=0$ and $m_{S}=\mp1$
manifolds, respectively. The single peak in the left or right transition
of the $^{14}$N multiplet verifies the operation of the CNOT gate,
which transfers population from states $|0\downarrow\rangle$ to $|-1\downarrow\rangle$,
or $|0\uparrow\rangle$ to $|1\uparrow\rangle$, respectively. We
use the amplitude ratio between the single peak and the corresponding
peak in the doublet to estimate the fidelity of the CNOT gate, and
obtained values of 0.99 and 0.95 for b) and d), respectively. 

If we use a transition-selective pulse to implement the CNOT gate,
its Rabi frequency may not exceed 20 kHz if the theoretical fidelity
should be at least $0.99$ in the absence of dephasing effects. For
such a low Rabi frequency, the duration of the $\pi$ pulse becomes
at least 25 $\mu$s, which is longer than the relaxation time $T_{2}^{*}\approx10$
$\mu$s. Dephasing would thus reduce the fidelity of such a gate operation
to $\sim0.6$.

\section{Offset dependence}

\subsection{Narrow frequency range}

The operations implemented by the $(\pi/2)_{x}-1/2|A|-(\pi/2)_{y}$
pulse sequence depend strongly on the MW carrier frequency
used to generate the pulses, as evidenced by the expression for $U$
in Eq. (\ref{UCNOTt}). We therefore measured this dependence in the
system consisting of the electron and $^{14}$N nuclear spins as a
function of the carrier frequency and measuring the resulting ground
state population. Applying the unitary of Eq. (\ref{UCNOTt}) to the
initial state (\ref{initials2}) yields ground state population 
\begin{equation}
P_{|0\rangle}=\frac{1}{2}-\frac{1}{6}\sin(\frac{\pi\nu}{A}).\label{Restheoy}
\end{equation}

\begin{figure}
\centering{}\includegraphics[width=1\columnwidth]{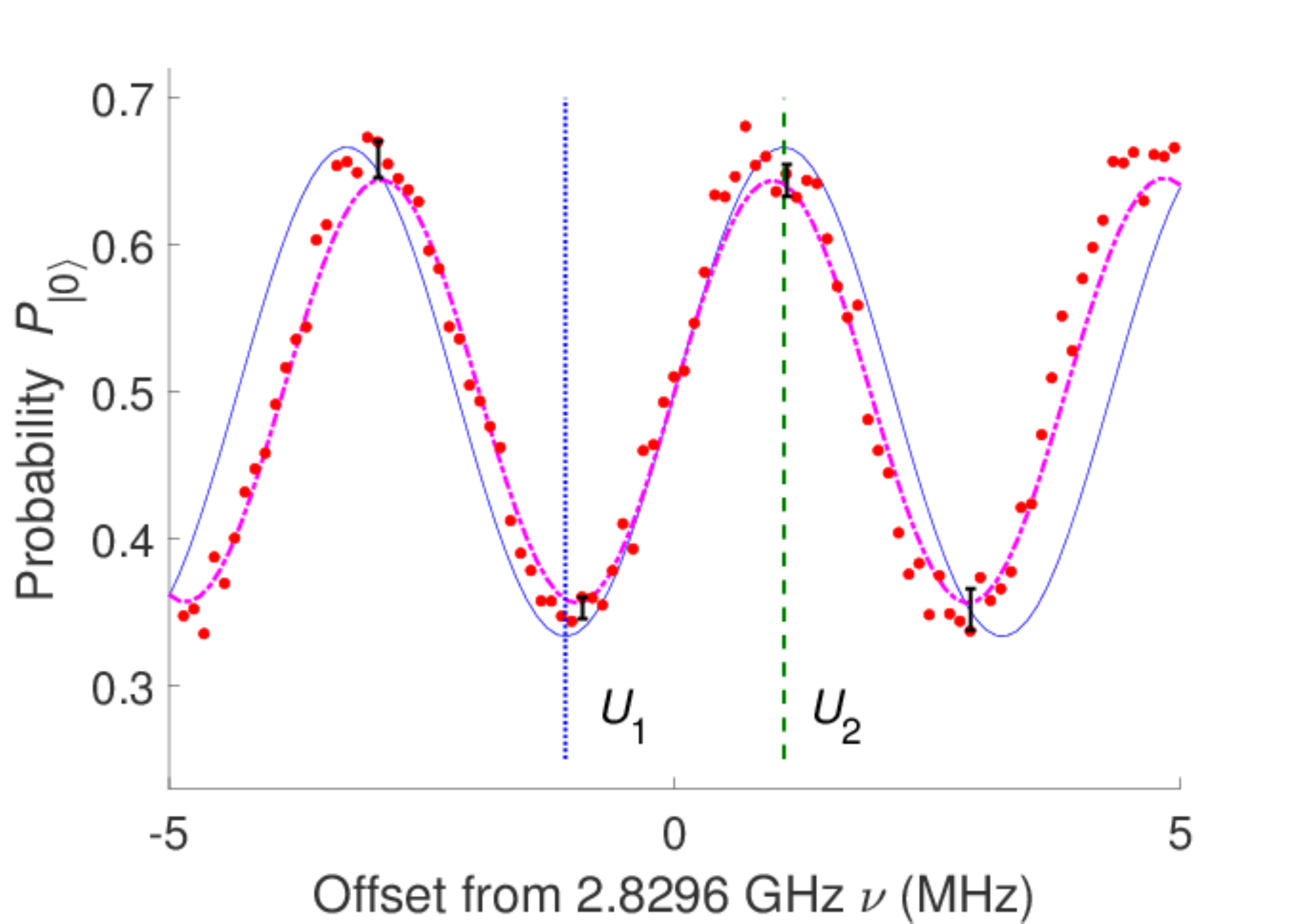}
\caption{(color online). Dependence of the operation $U$ defined by Eq. \eqref{UCNOTt} on
the MW carrier frequency. The filled circles represent the experimental
data, the full curve represents a simulation for ideal ($\delta$-function)
pulses, while the dash-dotted line was simulated with the pulse durations
of 21.6 $\mu$s used in the experiment. The blue dotted vertical line
indicates the offset used for implementing $U_{1}$ and the dark green
dashed line for $U_{2}$. The error bars represent the standard deviation
of the measured data points. \label{figCNOTemain}}
\end{figure}

Figure \ref{figCNOTemain} shows the results of this experiment: The
vertical axis represents the population of the $m_{S}=0$ state measured
by the second laser pulse as a function of the MW carrier frequency
$\nu$. The dotted and dashed vertical lines indicate the offsets
where the operation corresponds to $U_{1}$ and $U_{2}$, respectively.
We compare the experimental results (filled circles) to a simulation
of the experiment without dephasing effects, i.e. $T_{2}\rightarrow\infty$,
and using ideal pulses with infinite Rabi frequency of the MW pulses,
shown as the full curve. The dash-dotted curve in the figure shows the
result of a simulation for the actual Rabi frequency ($11.6$ MHz)
used in the experiment, and $T_{2}\rightarrow\infty$. It agrees well
with the experimental data, indicating that $T_{2}$ effects are negligible
in this experiment.

\subsection{Wide frequency range}

\label{spin1e} 
\begin{figure}
\centering{}\includegraphics[trim=1.2cm 0cm 1.7cm 0.5cm, width=8cm]{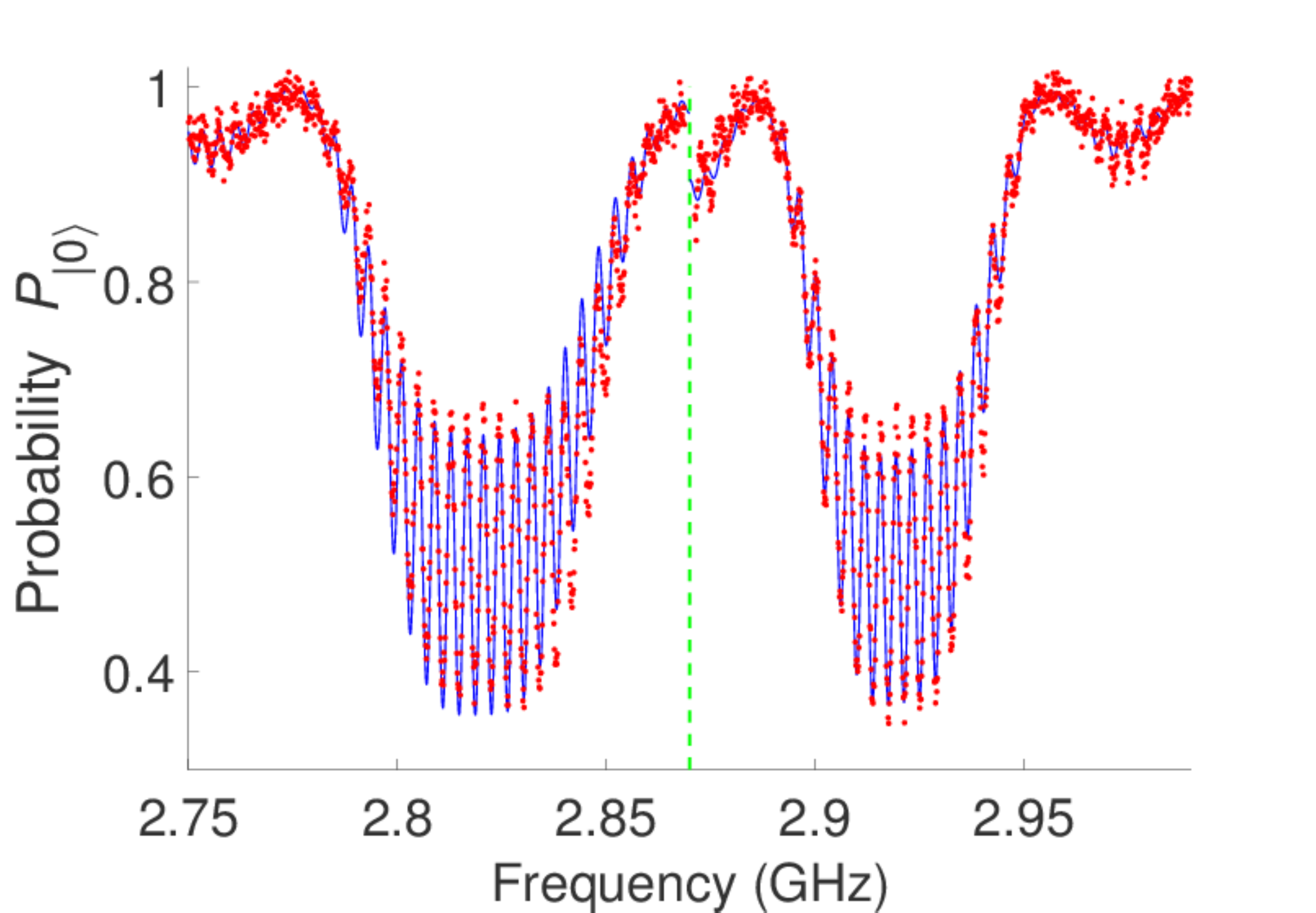}
\caption{(color online). Offset dependence of a CNOT gate with $^{14}$N as
the results in experiment and by simulation where the electron and
$^{14}$N spins are both treated as spin-1 systems, indicated by the
filled dots and the solid curves, respectively. The horizontal axis
shows the carrier frequency of the MW. We use two experiments by separating
the scan into two segments, indicated by the dashed line, since the
measured the Rabi frequencies at the transition frequencies are different.
\label{figCNOTlargeSpin1}}
\end{figure}

In addition to the data shown in Figure \ref{figCNOTemain}, we also
measured the offset dependence over the whole frequency range from
2.75 to 2.99 GHz, to obtain the full offset dependence. Figure \ref{figCNOTlargeSpin1}
shows the experimental results and compares them to a simulation that
treats the nuclear and the electronic spins as spin-1 systems. Since
the MW power at the sample varies as a function of frequency,
we performed the experiment in two parts and set the duration of the
$\pi/2$ pulses to 21.6 $\mu$s in the low-frequency part and to 27.5
$\mu$s in the high-frequency part. The two parts are separated by
the dashed vertical line in figure \ref{figCNOTlargeSpin1}. The agreement
between theory and experiment is best near 2.83 GHz and 2.93 GHz,
where the calibration for the pulse duration was performed.

\subsection{Transition-selective pulse}

\begin{figure}
\centering{}\includegraphics[width=1\columnwidth]{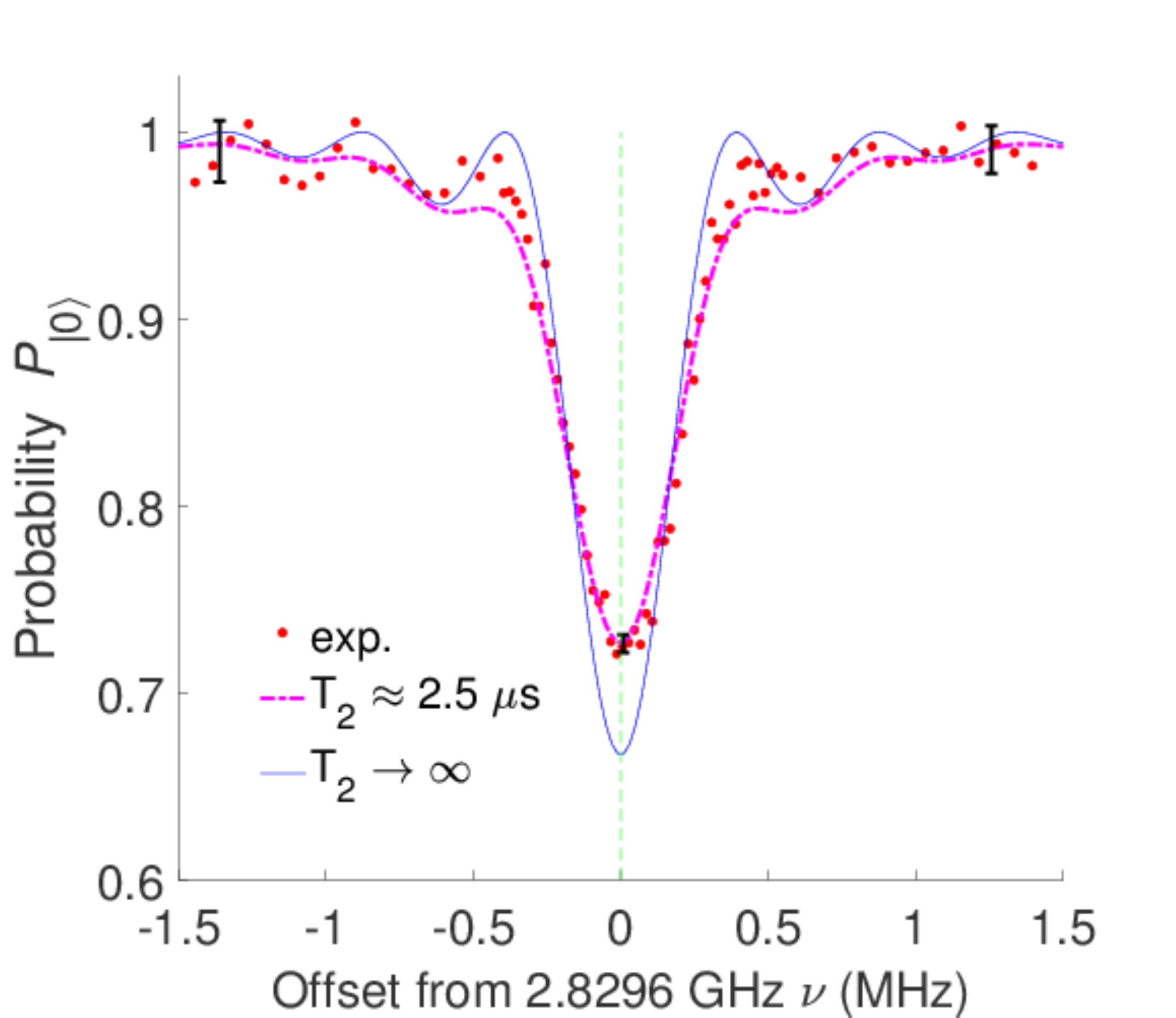}
\caption{(color online). Experimental results from a CNOT gate using a low-power pulse with
a Rabi frequency of $0.23$ MHz shown as circles. The dash-dotted
and full curves show numerical simulations for the cases of $T_{2}\approx2.5$
$\mu$s and $T_{2}\to\infty$. The dashed vertical line indicates
the offset for implementing $U_{2}$. \label{figCNOTemainb}}
\end{figure}

For comparison, we also implemented $U_{2}$ in the conventional way
\cite{PhysRevLett.93.130501,PhysRevLett.115.110502,nature17404,nature10401},
using one low power pulse with a Rabi frequency of $0.23$ MHz. Figure
\ref{figCNOTemainb} show the experimental and simulated results.
The comparison between the experimental and simulated data shows that
the $T_{2}$ effects cause $\sim17\%$ loss of the fidelity of the
gate, which is implemented at offset $0$.

\section{Conclusions}

The method for implementing CNOT gate based on transition-selective
pulse is an approximation \cite{quant-ph:9809045}, and the fidelity
is limited by the spectral resolution of the system. Our results show
that these limitations can be circumvented by using quantum gates
based on hard pulses and free precession periods. The resulting gate
duration is considerably shorter, which reduces the effects of relaxation.
In our experiments, we can implement a CNOT gate within $300$ ns
in the electron and nitrogen system, which is shorter than the pulse
obtained by optimal control ($\sim450$ ns) \cite{PhysRevLett.117.170501},
and much shorter than the transition-selective pulse ($~1.8\mathrm{\mu s}$)
\cite{PhysRevLett.115.110502}. The advantages were enhanced in the
electron and $^{13}$C system with the small coupling strength, e.g.,
about $150$ kHz, where the spectral resolution seriously limits the
fidelity of gates based on transition-selective pulses.

In conclusion, this new scheme for multi-qubit operations provides
significant improvements for quantum registers consisting of electronic
and nuclear spins. The benefits include reducing the gate time, and
improving the control in the system with more couplings. The gates
that we have chosen to implement here are important since they are
members of the sets of gates required for universal quantum computing.
Although we implemented the experiments in the NV center system, the method
is completely general and can be applied to many other systems.

\section{acknowledgement}

This work was supported by the DFG through grant SU 192/34-1.

 \bibliographystyle{apsrev}

\end{document}